\def\astroph{1}

\ifnum\astroph=0
\documentclass[12pt,preprint]{aastex}       
\usepackage{epsfig,amsmath,natbib}
\def\w{0.75}
\def\black{dashed black}
\def\grey{solid grey}
\graphicspath{{ApJ1/}}

\else
\documentclass{emulateapj}
\usepackage{apjfonts}
\usepackage{epsfig,amsmath,natbib}
\graphicspath{{arXiv/}}
\def\w{1}
\def\black{dashed red}
\def\grey{solid blue}
\fi

\newcommand{\vect}[1]{\ensuremath{\mbox{\boldmath$\mathrm{#1}$}}}
\newcommand{\vectdot}[1]{\ensuremath{\mbox{\boldmath$\dot{\mathrm{#1}}$}}}

\begin{document}

\shorttitle{GRB synthetic spectra} \shortauthors{Hededal and
Nordlund}

\title{Gamma-Ray Burst Synthetic Spectra from Collisionless Shock PIC Simulations}
\author{Christian Busk Hededal$^1$ and \AA{}ke Nordlund$^2$}
\affil{$^1$ Dark Cosmology Centre, the Niels Bohr Institute, $^2$
The Niels Bohr Institute, \\Juliane Maries Vej 30, 2100 K\o benhavn
\O , Denmark} \email{hededal@astro.ku.dk}

\begin{abstract}
The radiation from afterglows of gamma-ray bursts is generated in
the collisionless plasma shock interface between a relativistic
outflow and a quiescent circum-burst medium. The two main
ingredients responsible for the radiation are high-energy,
non-thermal electrons and a strong magnetic field. In this Letter we
present, for the first time, synthetic spectra extracted directly
from first principles particle-in-cell simulations of relativist
collisionless plasma shocks. The spectra are generated by a
numerical Fourier transformation of the electrical far-field from
each of a large number of particles, sampled directly from the
particle-in-cell simulations. Both the electromagnetic field and the
non-thermal particle acceleration are self-consistent products of
the Weibel two-stream instability. We find that the radiation
spectrum from a $\Gamma=15$ shock simulation show great resemblance
with observed GRB spectra---we compare specifically with that of GRB
000301C.
\end{abstract}


\section{Introduction}

Radiation from gamma-ray burst (GRB) afterglows is generally
believed to be synchrotron radiation. The main ingredients in this
type of radiation are a strong magnetic field and a power-law
distribution of high-energy electrons. Multi-wavelength observations
are more or less well fitted by hiding our ignorance about the
microphysical details of these ingredients in the parameters:
$\epsilon_B$ and $\epsilon_e$ describing the amount of the total
shock energy that is deposited in magnetic field and electrons
respectively, and $p$; the slope of the non-thermal electron
distribution, $N(\gamma)d\gamma\propto\gamma^{-p}$ where
$\gamma=[1-(v/c)^2]^{-1/2}$ is the relativistic Lorentz factor.
However, the true origin and nature of these two ingredients remain
unexplained.

 Regarding the
magnetic field, observations indicate that the typical value of the
equipartition parameter in afterglows is $\epsilon_B=0.0001-0.1$
\citep{bib:waxman1997a,bib:wijers1999,bib:Panaitescu+Kumar,bib:yost2003}.
This is many orders of magnitude larger than what can be achieved by
shock compression of the interstellar medium. Even if the field is
injected from the progenitor it is hard to maintain such a high
level of magnetic field in the shock region.
\cite{1999ApJ...511..852G} and \cite{bib:medvedevloeb} suggested
that the Weibel two-stream instability can generate a strong
turbulent magnetic field in the shock region \citep{bib:weibel}.
This has been confirmed by numerical simulations
\citep{bib:silva,bib:frederiksen2004,bib:nishikawa,bib:nishikawa2004,%
bib:hededal2004,bib:hededal2005,bib:hededal_thesis}. It has
furthermore been suggested that the radiation from non-thermal
particles in this turbulent magnetic field can be described as
'jitter-' or 'diffuse synchrotron' radiation
\citep{bib:Medvedev_jitter,bib:medvedev2005a,bib:fleishman2005}.

Non-thermal acceleration of particles is most commonly
described as Fermi acceleration \citep[e.g.][and references therein]%
{bib:Niemiec}. Fermi acceleration relies heavily on assumptions
about the nature of the magnetic field in the upstream/downstream
region. The Fermi acceleration mechanism also faces several
observational problems: E.g.\ in the Crab Nebula, the low energy
electrons have a power-law distribution spectrum $1.1 \le p \le
1.3$, much lower than the "universal" $p = 2.2-2.4$
\citep{bib:weiler}. Additionally, with Fermi acceleration, one
expects the presence of an X-ray halo around the shock, but such a
halo is not seen in Chandra observations of SN 1006
\citep{bib:long}. Recently \cite{bib:baring} found that particle
distribution functions (PDFs) inferred from GRB observations are in
contradistinction with standard acceleration mechanisms such as
diffusive Fermi acceleration. We stress that the Fermi acceleration
mechanism in principle may work, but that it has never been
explicitly demonstrated to dominate, or even to be significant, in
first principles simulations.

We are entering an era where very large simulations with
three-dimensional particle-in-cell (PIC) codes are becoming
computationally affordable. Such codes work from first principles
and can be considered as numerical experiments. 3D PIC experiments
of relativistic collisionless plasma shocks have been performed by
\citet{bib:silva,bib:frederiksen2004,bib:nishikawa,bib:nishikawa2004,bib:hededal2004,bib:hededal2005,bib:hededal_thesis}.
These simulations have shown that the two-stream-instability
generated magnetic field is highly turbulent and varies greatly
through the shock region. In the non-linear stage of the
instability, filaments merge and can become as strong as
$\epsilon_B\le 0.1$. Recently, \cite{bib:hededal2004} have further
shown that the Weibel two-stream instability also involves
non-thermal acceleration of the electrons. This new acceleration
mechanism differs from Fermi acceleration in the sense that it is
local and instantaneous. Electrons are trapped in the potential of
the generated current filaments and oscillate with strong
acceleration / re-acceleration. The numerical experiments have also
shown that the collisionless shock transition zone is at least of
the order hundreds of ion skin-depths
\citep{bib:frederiksen2004,bib:hededal2004,bib:hededal_thesis}. This
is considerably thicker than previously thought
\citep{bib:gruzinov2001}.

In this Letter we present a novel tool that allow us for the first
time to extract radiation spectra directly from PIC experiments. In
section \ref{sec:pic} we present the radiation tool , in section
\ref{sec:jit} we apply the tool to investigate the 3D jitter
radiation spectrum, and in section \ref{sec:exp} we present
synthetic radiation spectra from a 3D relativistic shock simulation
resembling a GRB afterglow collisionless shock. Finally, in Section
\ref{sec:discuss} we give a brief discussion on the results and
compare with observations.

\section{Synthetic spectra from PIC-code simulations}\label{sec:pic}
A particle-in-cell (PIC) code simulates plasma on a much more
fundamental level than MHD. The code works from first principles, by
solving the Maxwell equations with source terms for the
electromagnetic fields, together with the relativistic equation of
motion for a large number of charged particles. The electromagnetic
fields are discretized on a three-dimensional grid whereas the
particles are defined with continuous positions and momenta within
the grid. The relativistic PIC code that we have used is described
in \cite{bib:frederiksen2004} and \cite{bib:hededal2004}. Recently,
radiative cooling and synthetic spectra generation has been added to
the code \citep{bib:hededal_thesis}.

As a particle undergoes acceleration it will emit radiation. In the
PIC code we know this acceleration at each time step and hence we
can reduce the energy of the particles in accordance with the Larmor
radiation formula. The energy loss is highly dependent on the
particles Lorentz factor and for relativistic particles, most of the
radiated energy is beamed into a narrow cone with opening angle
$1/\gamma$ around the momentum vector.  Because of this relativistic
beaming, we take an approximative approach and subtract the momentum
lost to radiation along the velocity vector of the particle. We have
tested this implementation of radiative cooling against the
analytical result for a gyro-orbiting electron in a homogenous
magnetic field and find excellent agreement
\citep{bib:hededal_thesis}.

Given the detailed information of particle positions, velocities,
and accelerations it is possible to compute the emitted radiation
spectrum. To do this, we start by adopting the expression from
\cite{bib:jackson} for the retarded electric field from a charged
particle moving with instantaneous velocity
$\vect{\beta}=\vect{v}/c$ under acceleration
$\vectdot{\beta}=\vectdot{v}/c$,
\begin{equation}
\vect{E}=\frac{q}{4\pi\epsilon_0 c}\left[\frac{\vect{n} \times
\{(\vect{n}-\vect{\beta})\times\vectdot{\beta}\}}
{\left(1-\vect{n\cdot\beta}\right)^3
R}\right]_\mathrm{ret}\label{eq:e_rad_field}.
\end{equation}
Here, $q$ is the unit charge, $c$ is the speed of light,
$\epsilon_0$ is the vacuum electric permittivity,
 and
$R$ is the distance to an observer. $\vect{n}$ is a unit vector that
points from the particles retarded position towards the observer. In
Eq.\ \ref{eq:e_rad_field} we have neglected a "velocity term", which
is justified when the typical length scale of the emitter is
negligible compared to the distance to the observer
\citep{bib:jackson}. The radiation spectrum from an accelerated
charge is then found by Fourier transforming the Poynting flux based
on Eq.\ \ref{eq:e_rad_field} in the retarded time frame. This gives
us the energy $dW$ radiated per unit solid angle $d\Omega$ and unit
frequency $d\omega$
\begin{equation}
\frac{d^2W}{d\Omega d\omega}=\frac{\mu_0 c
q^2}{16\pi^3}\left|\int_{-\infty}^\infty
{\frac{\vect{n}\times[(\vect{n}-\vect{\beta})\times\vectdot{\beta}]}{(1-\vect{n\cdot\beta})^2}
\ e^{i\omega(t'-\vect{n\cdot r}_0(t')/c)}dt'}
\right|^2\label{eq:retard_fourier}.
\end{equation}
This expression can only be analytically integrated under constrains
on the morphology of the electromagnetic field. For example, in the
case of a relativistic electron moving in a homogenous magnetic
field with zero electric field, the integral reduces to the well
known synchrotron case \citep{bib:rybicki}. In the general
electromagnetic case, no analytical solution exists. However, from
the PIC experiments we readily have positions, velocities and
accelerations for the particles that define the plasma. With this
information, we can numerically integrate Eq.\
\ref{eq:retard_fourier} to obtain a synthetic radiation spectrum
from each of the particles in a plasma. Finally we can add the
spectrum of each particle into a total spectrum. Adding the spectrum
from each particle linearly is valid as long as the phase of each
contribution is completely uncorrelated to the others.

\begin{figure}[!ht]
\begin{center}
\epsfig{figure=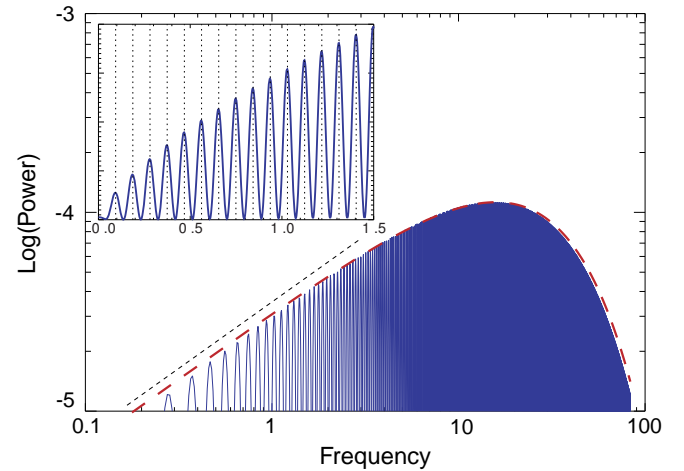,width=\w\linewidth}
\caption{The
spectrum obtained from a single particle gyrating in a magnetic
field from the code ({\it \grey}). The {\black} line shows
the theoretical synchrotron spectrum. The dashed line indicates the
power-law slope $2/3$. The inset in the upper left corner shows that
the spectrum consists of a discrete set of spikes that are integer
overtones of the gyrofrequency, in this case $\omega_B=0.094$
(arbitrary units) ({\it dotted vertical lines}). The particle has a
Lorentz factor $\gamma=5$. All numbers are in simulations units.}
\label{fig:synch_single}
\end{center}
\end{figure}

We have implemented Eq.\ \ref{eq:retard_fourier} and tested several
scenarios where an analytical solution exist. This includes
synchrotron radiation, bremsstrahlung, and wiggler/undulator
radiation. In all tests we find that our radiation tool can
effectively reproduce the theoretical expected solutions. Figure
\ref{fig:synch_single} shows the synthetic spectrum (\grey) of a
single relativistic electron gyrating in a homogenous magnetic
field. The envelope curve (\black) shows the analytical solution,
which shows good agreement with the simulations. The inset in Fig.\
\ref{fig:synch_single} shows that the peaks in the spectrum fall on
the relativistic gyro-frequency and its higher harmonics as
expected. For more details and tests, see \cite{bib:hededal_thesis}.

\section{Example: 3D Jitter Radiation}\label{sec:jit}
The PIC plasma simulation code, in combination with the radiation
generation module, provides us with a powerful tool for testing
various non-linear problems. In this section we examine the
so-called jitter- or diffuse synchrotron radiation spectrum
\citep{bib:Medvedev_jitter,bib:medvedev2005a,bib:fleishman2005}.
This type of radiation is generated by relativistic particles that
are deflected in a turbulent magnetic field, where the deflection
angle is comparable to the relativistic beaming angle.

\begin{figure}[!t]
\begin{center}
\epsfig{figure=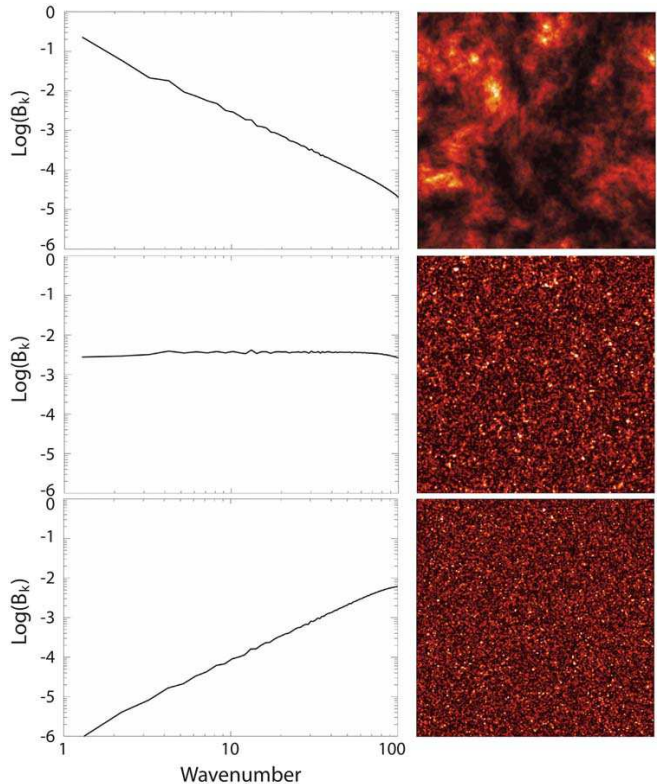,width=\w\linewidth}
\caption{{\it Left
column:} The Fourier structure of the turbulent magnetic field used
in the jitter radiation experiments. Here we show three different
values of the spectral power law index $\mu$: $\mu=-2$ (red noise,
{\it top}), $\mu=0$ (white noise, {\it middle}) and $\mu=2$ (blue
noise, {\it bottom}). {\it Right column:} Corresponding spatial
2D-slices of the magnetic field.}\label{fig:div_mu_cont}
\end{center}
\end{figure}

We setup PIC experiments with a periodic random magnetic field with
a power spectrum that follows a power-law distribution in the
Fourier domain $P_B(k)\propto k^{\mu}$ (Fig.\
\ref{fig:div_mu_cont}). In this field we use the PIC code to trace
an isotropic momentum distributions of electrons with a Lorentz
factor $\gamma=3$. The strength of the magnetic field is set such
that the maximum deflection of the electrons is smaller than the
relativistic beaming angle for all Fourier nodes.

\begin{figure}[!t]
\begin{center}
\epsfig{figure=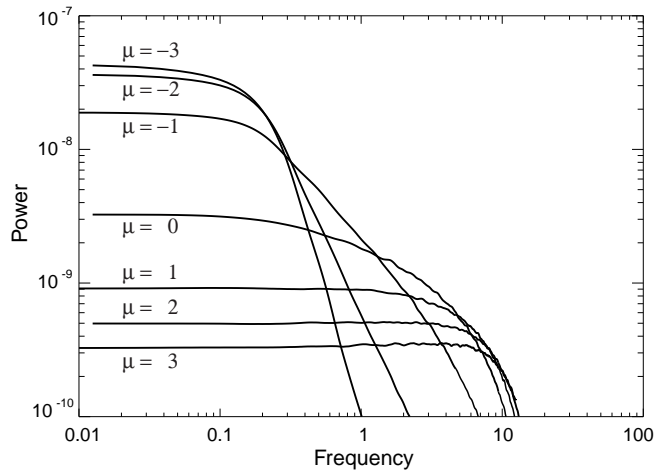,width=\w\linewidth}
\caption{The jitter
spectrum from a mono-energetic ensemble of electrons ($\gamma=3$)
moving in a turbulent magnetic field. The different graphs are for
simulations with different values of $\mu$.} \label{fig:div_mu}
\end{center}
\end{figure}

Fig.\ \ref{fig:div_mu} shows the resulting radiation spectrum for
different values of the slope $\mu$ of the magnetic Fourier
decomposition $P_B(k)\propto k^{\mu}$. For $\mu<0$, the high-energy
part of the spectrum follows a power-law with a slope
$\alpha\simeq\mu-1$ independent of the electron energy. For $\mu>0$,
the flat spectrum continues to higher frequencies than for $\mu<0$
and the spectrum has a hard cut-off at high frequencies. In all
cases, the low energy part is flat and quite similar to the case of
relativistic bremsstrahlung.

We observe that as the strength of the magnetic field is increased,
so that the typical deflection angle of the electrons becomes larger
than the relativistic beaming angle, the resulting spectrum
converges through the wiggler domain to the synchrotron case
\cite{bib:attwood}.

\section{Radiation from Collisionless Shock}\label{sec:exp}
Our goal is to obtain spectra from the PIC shock experiments that
may be compared directly with observations. Even though it is beyond
the scope of this Letter, this will eventually allow us to put
constraints on the conditions and physics of GRB afterglows such as
GRB environments and jet structure.

We use the numerical PIC experiment studied in
\citet{bib:hededal2004}. Two colliding plasma populations are
studied in the rest frame of one of the populations (downstream,
e.g.\ representing the shocked ISM medium at the head of a GRB jet).
In this frame of reference, a less dense population (upstream,
representing the unshocked interstellar medium) is continuously
injected at the left hand side boundary, with a relativistic
velocity corresponding to a Lorentz factor $\Gamma=15$. The two
populations initially differ in density by a factor of 3. We use a
computational box with $125\times125\times2000$ grid points and a
total of $8\times10^8$ particles. The ion rest-frame plasma
frequency in the downstream medium is $\omega_{pi}=0.075$, rendering
the box 150 ion skin depths long. The electron rest-frame plasma
frequency is $\omega_{pe}=0.3$ in order to resolve also the
microphysics of the electrons. Hence, the ion-to-electron mass ratio
is $m_i/m_e = 16$.

As the two plasma populations interpenetrate, we observe how the
Weibel two-stream instability is excited and current filaments are
formed
\citep{bib:weibel,bib:medvedevloeb,bib:silva,bib:frederiksen2004,bib:nishikawa,bib:nishikawa2004,bib:hededal2004}.
The current filaments induce a highly intermittent electromagnetic
field with a strength of up to 10\% of equipartition. The nature of
the magnetic field is turbulent, with a power spectrum that follows
a power-law in the Fourier domain \citep{bib:frederiksen2004}. We
note however, that unlike the random fields in section \ref{sec:jit}
there exists phase-correlations in the generated magnetic field that
give the field a more coherent structure. The instability also
directly leads to acceleration of electrons through a newly
discovered mechanism \citep{bib:hededal2004,bib:nishikawa2005}.
Electrons caught within the Debye cylinder surrounding the ion
current filaments are strongly and repeatedly accelerated and
decelerated. The acceleration is instantaneous and local, and thus
differs substantially from recursive acceleration mechanisms such as
Fermi acceleration. The resulting particle distribution function has
a non-thermal high-energy component, correlated with the
distribution of the generated electromagnetic filaments. We note in
passing that a fraction of the ions are scattered back into the
interstellar medium. Thus, ion Fermi acceleration may possibly take
place in collisionless shocks.

To obtain the total spectrum from the accelerated particles in the
Weibel-generated field, we have traced 20,000 particles from the PIC
simulation described above. Placing the observer somewhere along the
jet direction we have numerically integrated Eq.\
\ref{eq:retard_fourier} for each of these particles.

\begin{figure}[!t]
\begin{center}
\epsfig{figure=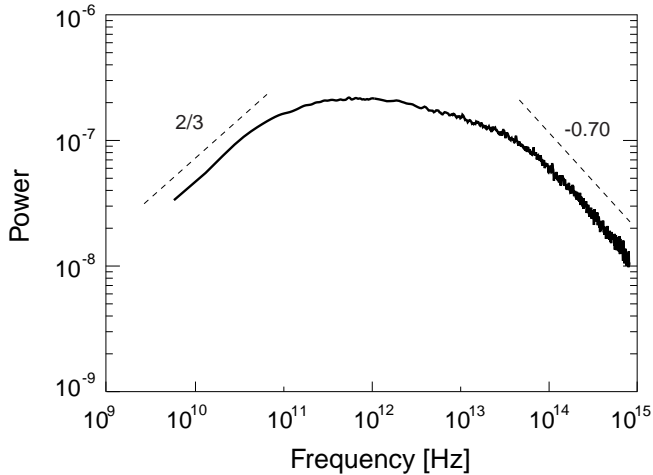,width=\w\linewidth} \caption{The
spectrum from a $\Gamma=15$ plasma shock propagating into an
ISM-like medium. The spectrum peaks in the far infrared. Below the
peak, we find a power-law segment with slope
$P(\omega)\propto\omega^{2/3}$. For frequencies above the peak
frequency, the spectrum continues into the near infrared/optical
band, following a power-law with slope
$P(\omega)\propto\omega^{-\beta}$, where $\beta=0.7$.}
\label{fig:spec_g15}
\end{center}
\end{figure}

Fig.\ \ref{fig:spec_g15} shows the radiated spectrum from the
collisionless shock simulations described above. Here, we have
rescaled the frequency into real space units and taken the
relativistic Doppler-shift into account. We find that the spectrum
peaks in the far infrared.  Below the peak, we find a power-law
segment with slope $P(\omega)\propto\omega^{2/3}$. This is
interesting because it is steeper than the limiting 1/3 slope for
synchrotron radiation, and may reconcile theory and observations on
the "line of death" issue \citep{bib:preece1998} of observed GRB
spectra. For frequencies above the peak frequency, the spectrum
continues into the near infrared / optical band, following a
power-law with slope $P(\omega)\propto\omega^{-\beta}$ with
$\beta\simeq0.7$.

From the power-law slope in the spectrum with ($\beta=0.7$), the
standard procedure commonly used in the GRB community would indicate
that the electrons have a power-law distribution, with slope $p=2.4$
(determined by solving $(p-1)/2=\beta$). The standard conclusion
would be that this is consistent with Fermi acceleration. However,
we strongly emphasize that the electrons that have produced the
spectrum in Fig.\ \ref{fig:spec_g15} are not accelerated by a
recursive mechanism such as Fermi acceleration. The electrons are
instantaneously accelerated and decelerated in the highly
complicated electric and magnetic field near the ion current
channels under the emission of strong radiation. This differs
substantially from the iterative acceleration in Fermi acceleration.
Furthermore, the paths of the accelerated shock electrons are closer
to a random walk than circular, which means that synchrotron
radiation is not an adequate formalism in the current context.

\section{Discussion and Conclusions}\label{sec:discuss}
The main source of information we have about gamma-ray bursts (GRBs)
is from multi-wavelength observations of GRB afterglows. The
radiation from GRB afterglows is generated in collisionless shocks
between the external plasma (ISM) and the GRB jet. This emphasizes
the crucial importance of a full understanding of the physics of and
the radiation mechanism in collisionless shocks. In the current
standard model, the radiation is assumed to be emitted as
synchrotron radiation from power-law distributed electrons in a
magnetic field. Following our lack of knowledge about the
plasma-physical details, it is common practice to parameterize this
ignorance with the dimensionless parameters $\epsilon_B$,
$\epsilon_e$ and $p$ ($\epsilon_B$ and $\epsilon_e$ express the
fraction of the total internal energy that is deposited in magnetic
field and electrons, respectively, and $p$ is the slope of the
supposedly power-law electron momentum distribution function).

Since collisionless shocks  are extremely non-linear and
self-interacting systems, the only way to study their microphysics
with any credence is through self-consistent, three-dimensional
relativistic particle simulations. Using this approach we have
previously explained the origin and nature of a strong
electromagnetic field in the shock \citep{bib:frederiksen2004} and
been able to identify a new non-thermal electron acceleration
mechanism \citep{bib:hededal2004}. These two ingredients are direct
and unavoidable consequences of the Weibel two-stream instability.

In this Letter we have developed a novel tool that allows us to
extract radiation spectra directly from the particle-in-cell
experiments. Rather than applying the synchrotron approximation
(assuming homogenous magnetic field and a single electron power-law
distributed population) we trace a large number of electrons in the
experiments, and calculate the exact radiation emitted from each
electrons. The tool has been thoroughly tested and successfully
reproduces spectra from synchrotron radiation, bremsstrahlung and
undulator/wiggler radiation from small-angle deflections
\citep{bib:hededal_thesis}.

The tool has been utilized for a parametric study of 3D jitter
radiation, where a population of electrons radiate as they move in a
turbulent magnetic field whose power spectrum follows a power-law in
the Fourier domain $P_B(k)\propto k^{\mu}$. We have examined how the
slope $\mu$ influence on the resulting radiation spectra. For
$\mu<0$, the high-energy part of the spectrum follows a power-law
with a slope $\alpha\simeq\mu-1$ independent of the electron energy.
For $\mu>0$, the flat spectrum continues to higher frequencies than
for $\mu<0$ and the spectrum has a hard cut-off for high
frequencies.

We have furthermore examined the radiation from a typical GRB
afterglow collisionless shock that propagates with $\Gamma=15$
through the interstellar medium. We find that the resulting
radiation spectrum  peaks  just below $10^{12}$Hz. Above this
frequency, the spectrum follows a power-law
 $F\propto\nu^{-\beta}$, with $\beta=0.7$. Below the peak frequency, the
spectrum follows a power law  $F\propto\nu^{\alpha}$ with
$\alpha\simeq2/3$. This is steeper than the standard synchrotron
value of $1/3$ and thus more compatible with observations. Both the
slope and the peak is consistent with observations (e.g.\
\cite{bib:panaitescu2001} who finds $\beta=0.67\pm0.04$, a peak at
$3\times10^{11}$Hz and $\Gamma>10$ for the afterglow of GRB 000301C
after five days).

Finally, we would like to stress the following interesting point:
 If one hides
all the real physical details of the magnetic field in the
dimensionless parameters $\epsilon_B$, $\epsilon_e$, and $p$ the
conclusion from standard synchrotron radiation theory would be that
the slope of the electron distribution $N(\gamma)\propto\gamma^{-p}$
is found by solving $-\beta=-(p-1)/2=-0.70\ \to\ p=2.4$ \citep[which
is consistent with standard analysis of the GRB 000301C afterglow;][]
{bib:panaitescu2001}. This would appear to be consistent with
Fermi acceleration and the supposedly universal $p\simeq2.2\pm0.2$.
But a closer look reveals that the acceleration mechanism is not
Fermi acceleration but rather a natural consequence of the Weibel
two-stream instability. The electrons are instantaneously
accelerated and decelerated in the highly complicated electric and
magnetic field near the ion current channels. Strong radiation is
produced in this process. The electron distribution behind the
radiation is a mixture of a thermal component for low energies and a
power-law component with $p=2.7$ for high energies
\citep{bib:hededal2004}. The paths of the electrons are more random
than circular and the electron distribution function varies
with shock depth, as does the magnetic topology and strength.

The outcome of the exercise is thus that it is indeed possible, from
first principles, to produce synthetic spectra very similar to the
ones that are observed in GRBs, and that the fit of such spectra
with the standard model assumptions may be more or less fortuitous.

CBH acknowledges support from the Instrument Centre for Danish
Astrophysics and from the University of Copenhagen for funding his
PhD grant. {\AA}N was supported by grant number 21-04-0503 from the
Danish Natural Science Research Council. Both authors thank Jacob
Trier Frederiksen and Troels Haugb\o{}lle for useful discussions and
comments. Computing resources were provided by the Danish Center for
Scientific Computing.

\bibliographystyle{apj}
\bibliography{radiation}

\end{document}